# Nitric Oxide Response to the April 2010 Electron Precipitation Event: Using WACCM and WACCM-D With and Without Medium-Energy Electrons

Christine Smith-Johnsen[1,2] 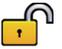, Daniel R. Marsh[3], Yvan Orsolini[2,4], Hilde Nesse Tyssøy[2], Koen Hendrickx[5], Marit Irene Sandanger[2], Linn-Kristine Glesnes Ødegaard[6], and Frode Stordal[1]

[1]Department of Geosciences, University of Oslo, Oslo, Norway, [2]Birkeland Centre for Space Science, University of Bergen, Bergen, Norway, [3]National Center for Atmospheric Research, Boulder, CO, USA, [4]Norwegian Institute for Air Research (NILU), Kjeller, Norway, [5]Department of Meteorology (MISU), Stockholm University, Stockholm, Sweden, [6]Royal Norwegian Naval Academy, Bergen, Norway

**Abstract** Energetic electrons from the magnetosphere deposit their energy in the atmosphere and lead to production of nitric oxide (NO) in the mesosphere and lower thermosphere. We study the atmospheric NO response to a geomagnetic storm in April 2010 with WACCM (Whole Atmosphere Community Climate Model). Modeled NO is compared to observations by Solar Occultation For Ice Experiment/Aeronomy of Ice in the Mesosphere at 72–82°S latitudes. We investigate the modeled NOs sensitivity to changes in energy and chemistry. The electron energy model input is either a parameterization of auroral electrons or a full range energy spectrum (1–750 keV) from National Oceanic and Atmospheric Administration/Polar Orbiting Environmental Satellites and European Organisation for the Exploitation of Meteorological Satellites/Meteorological Operational satellites. To study the importance of ion chemistry for the production of NO, WACCM-D, which has more complex ion chemistry, is used. Both standard WACCM and WACCM-D underestimate the storm time NO increase in the main production region (90–110 km), using both electron energy inputs. At and below 80 km, including medium-energy electrons (>30 keV) is important both for NO directly produced at this altitude region and for NO transported from other regions (indirect effect). By using WACCM-D the direct NO production is improved, while the indirect effects on NO suffer from the downward propagating deficiency above. In conclusion, both a full range energy spectrum and ion chemistry is needed throughout the mesosphere and lower thermosphere region to increase the direct and indirect contribution from electrons on NO.

## 1. Introduction

Energetic particle precipitation (EPP) ionizes the neutral atmosphere, leading to the formation $NO_x$ (N, NO, and $NO_2$) and $HO_x$ (H, OH, and $HO_2$) in the mesosphere and lower thermosphere (MLT; see review by Sinnhuber et al., 2012, and references therein). These chemically reactive species can catalyze ozone loss in the mesosphere and stratosphere (Brasseur & Solomon, 2003; Funke et al., 2011; Fytterer et al., 2015; Jackman et al., 2008, 2001; Verronen et al., 2006), which in turn has significant implications for the energy budget of the atmosphere.

The energetic particles are mostly protons and electrons, and their atmospheric influence is determined by their energy and source region (Vitt & Jackmann, 1996). Protons come directly from the Sun in large solar particle events (SPEs; Reames, 1999), which occur rarely but have a demonstrated impact on atmospheric ozone and other trace species (e.g., Funke et al., 2011; Jackman et al., 2001; Verronen et al., 2006). Auroral electrons originate in the Earth's plasma sheet and are accelerated to energies of 1–30 keV during auroral substorms (Fang et al., 2008). Medium-energy electrons (MEEs; >30 keV) precipitate from the terrestrial radiation belts, where they have been accelerated to energies of a few hundred kiloelectron volts during geomagnetic storms (Horne et al., 2009). When the electrons and protons reach the atmosphere, they can lead to production of $NO_x$ at a specific altitude level (called the direct effect of EPP) or influence other altitude regions when the long-lived $NO_x$ is vertically transported (the indirect effects of EPP; Randall et al., 2007).

In an observation based study by Smith-Johnsen et al. (2017), an energetic electron precipitation (EEP) event in April 2010 has been studied using a full range energy spectrum (FRES) covering 1–750 keV (corresponding to electrons depositing their energy at altitudes 60–120 km). FRES is derived from measurements by both the Total Energy Detector (TED) and Medium Energy Proton and Electron Detector (MEPED) on National Oceanic and Atmospheric Administration's (NOAA's) Polar Orbiting Environmental Satellite (POES). They showed that the NO direct effect can be detected down to 60 km during this event. The indirect effect, bringing down elevated NO from the middle or upper mesosphere, contributes strongly to the variability at these altitudes as well. In this paper we focus on simulating the atmospheric response to this EEP event in the MLT region (60–130 km).

In order to reproduce the effect of EEP in current chemistry-climate general circulation models, the energy input, chemical productions and losses, and transport need to be properly represented. In this study the Whole Atmosphere Community Climate Model (WACCM) is used (Marsh et al., 2013). We compare different model runs with the aim to disentangle the impact on the NO distribution from these different modeling aspects during the EEP event in April 2010. In WACCM (version 4), the standard EEP parametrization only includes auroral electrons (with a characteristic energy of 2 keV, restricting to energies <30 keV). We compare simulations with this auroral parameterization to simulations using the FRES estimated by Smith-Johnsen et al. (2017), in order to investigate WACCM's response to the different energy inputs and the importance of including MEE. We also carry out simulations with WACCM-D, a variant of WACCM with a more advanced ion chemistry scheme in the *D* region (Verronen et al., 2016), to investigate the sensitivity of modeled NO to ion and neutral chemistry.

This paper continues with section 2, describing the satellite observations used as model input and as model validation, the models WACCM and WACCM-D, and the auroral energy and full range electron energy input. In section 3 we describe the event with geomagnetic indices. Further, in section 4, modeled NO from the different runs is compared to observations by the Solar Occultation For Ice Experiment (SOFIE) instrument on National Aeronautics and Space Administration (NASA)'s Aeronomy of Ice in the Mesosphere (AIM) satellite. Finally, in the section 5 we discuss WACCM's sensitivity to energy input, to the chemistry schemes, and the influence of dynamics. Section 6 summarizes our findings.

## 2. Data and Methods
### 2.1. Whole Atmosphere Community Climate Model

WACCM is the National Center for Atmospheric Research's global chemistry climate model and is part of Community Earth System Model (Hurrell et al., 2013). The model version used is WACCM 4 (Marsh et al., 2013), which extends vertically from the ground to $5.9 \times 10^{-6}$ hPa (∼140-km geometric height), with 88 pressure levels and horizontal resolution of 1.9° latitude by 2.5° longitude. The Specified Dynamics version of WACCM (denoted SD-WACCM) is nudged with reanalysis data from NASA Global Modeling and Assimilation Office's Modern-Era Retrospective Analysis for Research and Applications (Rienecker et al., 2011) by the method described in Kunz et al. (2011), from the surface up to ∼50 km, with a transition region from ∼50 to 60 km, and is free running above ∼60 km. All our simulations are run with an enhanced eddy diffusion, with the Prandtl number halved, to improve the distribution of species with sharp vertical gradients in the MLT (Garcia et al., 2014), and NO in particular (Orsolini et al., 2017).

The parameterization of the electron energy deposition from auroral electrons in WACCM uses the planetary geomagnetic index *Kp* as input (Bartels, 1949; shown schematically in Figure 1, upper panels). *Kp* is a proxy for the global level of geomagnetic activity and is in WACCM used to calculate the hemispheric power (HP). HP is the estimated power deposited in the polar regions by EEP, and WACCM uses the relation of Zhang and Paxton (2008a) for *Kp* < 7:

$$\text{HP [GW]} = 16.82 \times e^{0.32 \times Kp} - 4.86; \qquad (1)$$

and the following linear relationship for *Kp* levels greater than 7:

$$\text{HP [GW]} = 153.13 + \frac{Kp - 7}{9 - 7} \times (300 - 153.13); \qquad (2)$$

This energy input is distributed in an auroral oval as described by Roble and Ridley (1987). From the HP the energy flux is also calculated, and it is assumed the auroral electrons have a Maxwellian energy distribution with a fixed characteristic energy of 2 keV (see Figure 1b), restricting to energies <30 keV. The energy deposition as a function of altitude is calculated as in Roble and Ridley (1987), which results in a profile that always peaks at ∼110 km (see Figure 1c or Figure 2b). The electron energy deposition ionizes the thermosphere and leads to the production of the major ions ($O^+$, $O_2^+$, $N^+$, $N_2^+$, and $NO^+$) and electrons in the *E* region (>90 km). From these, exited nitrogen $N(^2D)$ can be created, and then NO, through chemical reactions in the *E* region.

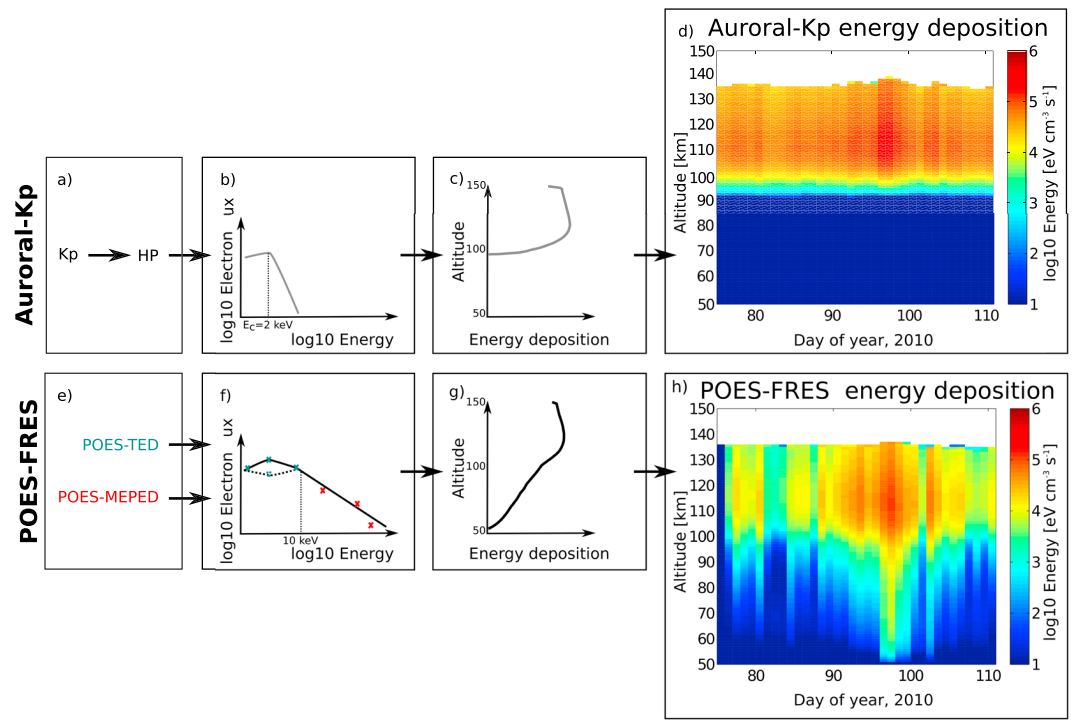

**Figure 1.** (a) The auroral-*Kp* parametrization for energetic electron precipitation (EEP) in Whole Atmosphere Community Climate Model version 4 is calculated from the geomagnetic *Kp* index, through hemispheric power based on Zhang and Paxton (2008b). (b) Then the electron flux per energy is calculated as a Maxwellian distribution with a fixed characteristic energy of 2 keV (Roble & Ridley, 1987). (c) The energy deposition per altitude is based on Roble and Ridley (1987). (d) Time and altitude distribution of auroral-*Kp* energy deposition (daily and hemispheric mean for latitudes >50°S) for the April 2010 EEP event. (e) Three hours of satellites passes from POES are interpolated into a polar map, one for each energy channel on TED and MEPED. (f) The three electron channels from TED (blue crosses) are fitted to a Maxwellian or Exponential distribution (for energies <10 keV) and the three MEPED channels (red crosses) to a power law for each latitude/longitude grid box (energies >10 keV). (g) The continuous energy range is converted to energy deposition profile following Rees (1989). (h) Time and altitude distribution of POES-FRES energy deposition (daily and hemispheric mean for latitudes >50°S) for the April 2010 EEP event. POES = Polar Orbiting Environmental Satellites; FRES = full range energy spectrum; TED = Total Energy Detector; MEPED = Medium Energy Proton and Electron Detector.

WACCM-D is a new version of WACCM with a more detailed ion chemistry relevant to the ionospheric *D* region (∼50–90 km). The extra chemical scheme is based on a simplification of the Sodankylä Ion Chemistry one-dimensional model (Verronen et al., 2016), with only the most important chemical reactions included based on their effects on the *D* region (Verronen & Lehmann, 2013). The aim is to better reproduce the observed impact of EPP in the *D* region. Where standard WACCM includes the five major ions in the *E* region, WACCM-D includes these five both in the *E* and *D* regions and has in addition 20 positive ions and 21 negative and 307 ion in the *D* region. This allows for NO production by EEP the same way as in the *E* region and also through multiple cluster ion-ion recombinations or by positive ion reactions in the *D* region. The ion chemistry is not limited to the *D* region but is included in WACCM-D over the whole altitude range. The positive ion clusters and negative ions are, however, less abundant above 90 km, so their effect becomes less important in the *E* region and the standard WACCM *E* region chemistry is still dominating here (Verronen et al., 2016).

### 2.2. AIM-SOFIE

The SOFIE instrument on board NASA's AIM satellite has been operational from May 2007 to present and measures trace gases in the polar middle atmosphere. AIM has a polar, Sun-synchronous orbit with a period of 96 min, so 15 orbits are performed each day. Vertical profiles of NO are retrieved two times per orbit, during sunset (Northern Hemisphere) and sunrise (Southern Hemisphere; Gordley et al., 2009). The NO abundance is retrieved from 30 up to 150 km, with a vertical resolution of 2 km. At the time of the April event in 2010, SOFIE's sampling of the Northern Hemisphere is outside the auroral region. In the Southern Hemisphere, SOFIE measures at geographical latitudes 72–82°S, which is both within and outside of the auroral region. Hence, we restrict our study to the Southern Hemisphere in the comparison with WACCM.

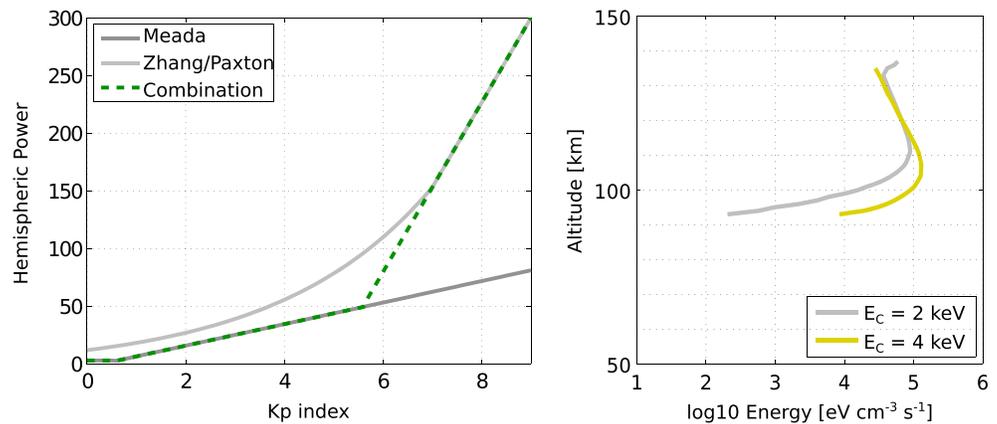

**Figure 2.** (left) Relation between *Kp* index and hemispheric power in the standard auroral energy input. The Zhang and Paxton (2008b) relation (light gray line) is now used in WACCM, while earlier WACCM versions used Maeda et al., 1989 (1989; dark gray line). We test a combination of the two (green dashed line). (right) Altitude distribution of energy deposition from the standard auroral energy deposition. Characteristic energy in the Maxwellian energy distribution is normally 2 keV (gray line). We test 4 keV (yellow line) to shift the energy further down into the atmosphere.

### 2.3. POES—TED and MEPED

The POES series by NOAA and European Organisation for the Exploitation of Meteorological Satellites are orbiting at 850 km with a period of 100 min. Six satellites were operating in the period of interest in April 2010, enabling us to construct global maps of the precipitating electrons with high spatial and temporal resolution (see description in Smith-Johnsen et al., 2017). The six spacecraft are equipped with two types of electron detectors, measuring low-energy electron and MEE with the instruments TED and MEPED, respectively. TED measures differential electron fluxes in the energy range 0.1–20 keV, MEPED measures high-energy integral electron fluxes at >43, >114, and >292 keV (Ødegaard et al., 2017). Both TED and MEPED have two telescopes, pointed in two directions. Here only the 0° telescopes have been used, which will provide a lower limit of the electron precipitation (Nesse Tyssøy et al., 2016). The MEPED electron detectors are known to be contaminated by low-energy protons (Evans & Greer, 2000; Yando et al., 2011). In order to correct for false counts, we use the proton flux measurement from the proton telescopes as described in Nesse Tyssøy et al. (2016) after the proton energy limits have been calibrated due to detector degradation (Ødegaard et al., 2016; Sandanger et al., 2015). We also account for relativistic (>750 keV) electrons contaminating the proton detector (Nesse Tyssøy et al., 2016; Ødegaard et al., 2017).

By combining the measurements from TED and MEPED, we are able to construct global maps of deposited energy in the range 1–750 keV, which can be used as input for the electron energy deposition in WACCM. This FRES covers both auroral and MEE energies. The six POES spacecraft will in a 3-hr period pass 12 times over the polar cap. For each energy channel, the passes are interpolated in corrected geomagnetic coordinates and then converted to geographical coordinates. To construct a continuous energy spectrum, the three TED channels are fitted to an exponential or a Maxwellian spectrum (depending on the ratio between the two first channels) and the MEPED channels to a power law fit (shown schematically in Figure 1, lower panels). The energy spectrum is converted to energy deposited as a function of altitude, using the cosine dependent Isotropic over the downward hemisphere of Rees (1989; see Figure 1g). Using a climatological atmospheric background will lead to some uncertainties in the calculated energy. The result is an energy deposition with 3-hr time resolution, 1-km altitude resolution (see Figure 1h), and a geographical resolution of 4° latitude times 10° longitude.

**Table 1**
*Overview of the Model Runs Performed*

| Model run | Electron energy | NO, *E* region | NO, *D* region |
|---|---|---|---|
| Standard | | | |
| Auroral-HP | Auroral (*Kp*) | *E* region ions | no energy deposited |
| Auroral-EC | | | |
| POES-FRES | Full range (POES) | *E* region ions | neutrals |
| POES-FRES D | Full range (POES) | *E* region ions | *D* region ions |

*Note.* The first column indicates the name of the model run. The electron energy input to the model is specified in the second column, where the parenthesis shows what the input is based on. How the energy deposition is transformed to NO production in the *E* region is listed in column three, and NO production in the *D* region is listed in column four.

### 2.4. Modified WACCM Runs

Five different model runs are performed. They are summarized in Table 1 below and will be describe further here. The first run is just standard WACCM with the original *Kp*-based parameterization of auroral electron precipitation.

In the second run, we look further at the HP relation and its influence on the EEP production of NO. HP in WACCM 4 is related to *Kp* as described by Zhang and Paxton (2008b; light gray in the left panel of Figure 2). In previous versions on WACCM, the relation by Maeda et al. (1989) was used (dark gray line). To test WACCM's sensitivity to the geomagnetic activity level, we have tested a combination of these parameterizations (green dashed line), where the Maeda relation is used for low *Kp* and the Zhang/Paxton relation for high *Kp*.

We also do a model run where we change the characteristic energy in the Maxwellian spectrum from 2 to 4 keV, to see how the simulated NO production is influenced by a production deeper in the atmosphere. The peak of the input energy is now shifted from ∼110- down to ∼105-km altitude (see the right panel in Figure 2).

In the fourth run, we replace the *Kp*-based parameterization of auroral electrons with FRES. The FRES data set covers both auroral and MEEs and is based on observations from the POES satellites as described above. In the standard run, EEP leads to production of *E* region ions, and from the ions $N(^2D)$ is formed and then NO. Running WACCM with FRES, NO will be produced also below 90 km, where no ions are present. WACCM's production of $N(^2D)$ will now be directly parameterized from the energy from the incoming electrons. It is assumed that the energy needed to produce one ion-pair is 35 eV (Porter et al., 1976), and each ion pair produces 1.25 nitrogen atoms. The amount of *N* produced by EEP is as follows:

$$N\,[\mathrm{cm}^{-3}/\mathrm{s}^{-1}] = \frac{\text{energy deposition}\,[\mathrm{eV}\cdot\mathrm{cm}^{-3}\cdot\mathrm{s}^{-1}]}{35\,[\mathrm{eV/ion\ pair}]} * 1.25\,[1/\mathrm{ion\ pair}] \quad (3)$$

Of this, 0.55 is assumed to be in the form $N(^4S)$ and 0.7 is $N(^2D)$ (Jackman et al., 2005). In the *E* region the ions will still be present, and in the *D* region NO can be produced through neutral reactions with $O_2$ or with OH.

The last run is WACCM-D with FRES input. Also here the relation in equation (3) is used to convert energy to production of $N(^2D)$, though with the ratio of 0.502 $N(^4S)$ and 0.583 $N(^2D)$. After the initial branching, the *D* region ions will lead to additional production of NO.

In the following we will present and evaluate the NO volume mixing ratio (VMR) variability seen in WACCM for the April 2010 period. To explore the sensitivity to specification of electrons fluxes, we apply these four different energy inputs. The *Kp*-based parameterization of the auroral electrons will be referred to as the Standard run. Then the two runs with modifications to the auroral energy deposition: The HP is changed (referred to as the Auroral-HP run), and the characteristic energy run is increased (auroral-EC run). When the auroral energy is replaced completely with the FRES from POES, the run is called POES-FRES. To test WACCM's sensitivity to *D* region ion chemistry, WACCM-D is used. The influence of *D* region ions is most efficient when energy is deposited in the *D* region, so WACCM-D is only run with the FRES input. This results in five different model runs, as summarized in Table 1.

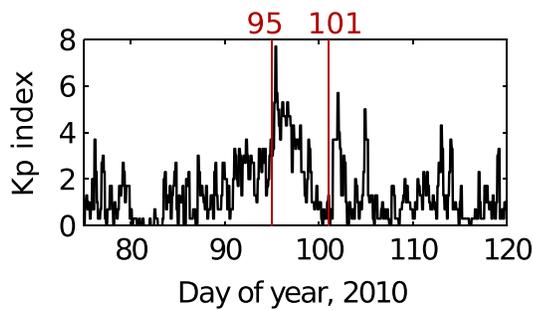

**Figure 3.** The *Kp* index (3-hr time resolution) for the geomagnetic storm period in April 2010. The two coronal mass ejections arrive at days of year 95 and 101.

## 3. April 2010 Event Description

The time period investigated in this study is April 2010. It includes a relativistic electron precipitation event caused by a combination of a high-speed solar wind stream and a subsequent corotating interaction region and two coronal mass ejections (CMEs) on the Sun. The CMEs arrive according to Richardson and Cane (2017) on 5 and 11 April 2010 (days of year [DOYs] 95 and 101). The highest *Kp* occurs at DOY 96 and will be referred to as the storm peak. The weeks before the storm period are characterized by geomagnetically quiet conditions, making the period of interest well isolated (see Figure 3 for *Kp* index). Smith-Johnsen et al. (2017) found evidence of direct production of NO down to 60 km during the April 2010 EEP event. They also pointed out that MEEs contributed significantly to the indirect effect at these altitudes.

## 4. Results

The aim of this paper is to show the sensitivity of simulated NO by WACCM to the choices of electron energy input and to choice of ion chemistry. Five model runs have been performed (see Table 1) and will be compared to NO observations from SOFIE.

### 4.1. Observed Nitric Oxide

The NO abundance measured by SOFIE during the April 2010 period has previously been reported by Smith-Johnsen et al. (2017) and is here illustrated in terms of VMR for different altitude regions by the black line in Figure 4. The main NO layer is typically found at 90–115 km in terms of number density, while NO VMR increases with higher altitudes toward the model top (at about 140 km).

The observed NO VMR starts to increase at DOY 92 at 120- and 130-km altitude. At the arrival of the CMEs (at DOYs 95 and 101), there is a reduction of NO lasting for about 2–3 days. After the CME arrivals the NO VMR increases again, reaching 200% at DOYs 97–98 and almost a 100% increase at DOYs 103–104 compared to the prestorm level.

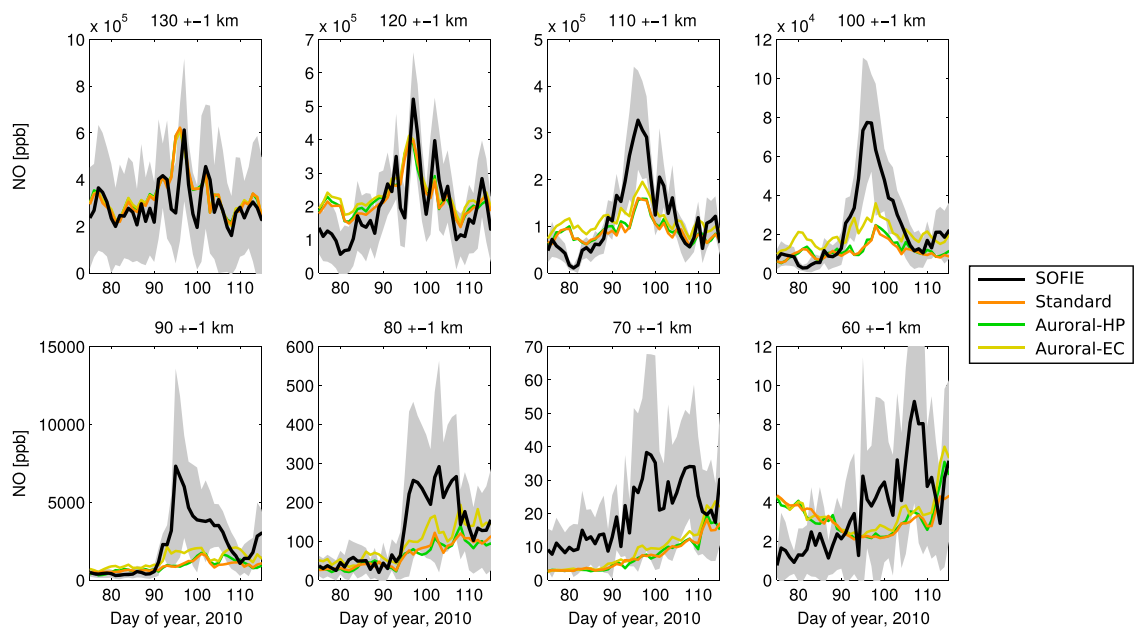

**Figure 4.** Time evolution of NO volume mixing ratio (VMR; ppb) for eight different altitude regions (daily mean, altitude mean ±1 km, and zonal mean of latitude band where Aeronomy of Ice in the Mesosphere-SOFIE is measuring). The black line represents SOFIE observations of NO VMR, where the observational spread is shown as the gray shaded region. The orange line represents the NO VMR from the Standard Whole Atmosphere Community Climate Model run, the green line represents auroral-hemispheric power run, and the yellow line from the auroral-EC run. The two CMEs arrive at days of year (DOY) 95 and 101. DOY 75 is 16 March, and DOY 120 is 30 April. EC = characteristic energy; SOFIE = Solar Occultation For Ice Experiment; HP = hemispheric power.

The relative changes associated with the geomagnetic storms are largest at 90- to 110-km altitude where it increases 300–400% after the first CME at DOYs 96–99. A new peak in the NO VMR occurs on DOYs 102–104, about a day after the second CME reaches Earth. The anomalously high NO VMR does not persist longer than 4 days after the CME arrival at altitudes between 100 and 130 km.

At and below 90 km, the NO VMR enhancement lingers longer. Here an increase is found for all altitudes at DOY 96, 1 day after the first CME arrives indicating direct NO production by EEP (Smith-Johnsen et al., 2017). At 90 km, in the aftermath of the events, the NO VMR reaches a plateau before decreasing after the second CME arrival. A broad peak is also characteristic for 80- and 70-km altitude, whereas the 60-km NO VMR appears to gradually increase after the first direct enhancement. For these altitudes, the maximum values are reached a day to a few days after the CME arrives. The lifetime of NO is constant through out this altitude range, so the gradual NO VMR increase in the aftermath of the CMEs is consistent with the NO being transported down after being produced at higher altitudes.

### 4.2. Modeled Nitric Oxide, With Different Auroral Electron Forcings

For the comparison with SOFIE, the modeled NO is taken from the same latitude and longitude as SOFIE is measuring. The 15 daily passes over the Southern Hemisphere are then averaged into a daily mean for both SOFIE and WACCM. The daily averaged NO VMR from WACCM, after collocation with SOFIE, is displayed as the orange line in Figure 4, for the Standard run with the original auroral electron energy deposition.

Comparing the Standard WACCM run with the SOFIE NO observation (orange and black lines in Figure 4, respectively), we find that, at 130 km, the modeled NO VMR is well in line with the observations, both during the quiet prestorm period and in the main phase of the event. Note that the model does not reproduce the observed NO decrease associate with the first CME. At 120 km, the model predicts the NO VMR level in the main phase of the event but overestimates the NO VMR in the geomagnetic quiet periods before. The overestimation of NO VMR during the prestorm period is also evident at 110 km, but here the modeled NO VMR also shows a deficit of a factor of 3–4 compared to the observations in the main phase of the storm. At 100 and 90 km, the storm time NO deficit becomes even more pronounced. Based on the Standard auroral electron energy deposition in Figure 1, we do not expect to see direct NO production at and below 90 km. We do, however, see an increase 3, 5, 8, and 10 days after the onset, at 90, 80, 70, and 60 km, respectively, which is due to NO produced at higher altitudes that has been transported to these lower altitudes. This would correspond to an effective transport rate of about 4 km/day assuming that the source region is at 100 km. The deficit found below 90 km is hence due to both lack of electron energy deposition input and the deficit found in the NO production at 100 km. During the geomagnetic quiet NO VMR background levels for the two lowermost altitudes at 70 and 60 km, the Standard run largely underestimates and overestimates NO VMR, respectively. In particular, at 60 km, the model shows a decrease with time before the event, while this is not seen in the observations.

To investigate the sensitivity of the modeled NO to the auroral electron energy input, we modified the Standard parameterization. We note that the modeled NO is too high during quiet times and too low during active conditions. By changing the HP and how it is related to the *Kp* index, we can alter the sensitivity to geomagnetic activity level. By increasing the characteristic energy in the Maxwellian energy distribution, we can shift the altitude of the peak electron energy input deeper into the atmosphere. In Figure 4 we show the time variation of NO from these two sensitivity runs (labeled auroral-HP and auroral-EC). There is little change in NO between 60 and 80 km for both modified runs. However, the bigger change is with auroral-EC, which improves the model/observation correspondence at 80–110 km—likely due to more in situ production at that height. The lower HP for quiet times does not lower the background level of NO and shows very little change compared to the Standard run.

We conclude that improving the existing auroral parameterization in this way is not sufficient to get the model to produce similar levels of NO as the observations show, especially at 110 km and below. The next section investigates the sensitivity to adding MEE, by replacing the auroral electron energy deposition with the FRES energy deposition based on electron flux observations from the POES satellites.

### 4.3. Modeled Nitric Oxide, With MEEs

To account for the lack of direct production throughout the mesosphere, we replace the Standard auroral electron energy deposition with POES-FRES, covering both auroral electrons and MEE (1–750 keV). The NO VMR modeled by WACCM with POES-FRES as energy input is shown as the red line in Figure 5. As was the case for the Standard run, the modeled NO using POES-FRES provides a good estimate at 120 and 130 km. POES-FRES is better at reproducing the prestorm NO VMR compared to the Standard run at 120 and 110 km but provides, however, lower NO production at these altitudes during the main peak of the storm. The largest deficit compared to the observations is found in the 90- to 110-km region. In contrast to the Standard run though, the NO VMR from POES-FRES peaks on DOY 96 in line with the energy deposition and the NO VMR observations, although largely underestimated. At 90 km, no energy is deposited from the Standard auroral input, while the POES-FRES inputs ~$10^3$ eV·cm$^{-3}$·s$^{-1}$ during the quiet prestorm period and ~$10^4$ eV·cm$^{-3}$·s$^{-1}$ during the storm peak on DOY 96. Despite all this extra energy input, the NO VMR is higher at quiet times in the Standard run compared to the POES-FRES run. During the storm peak at DOY 96, the NO VMR from the two model runs is similar. In the aftermath of the storm, the NO VMR found in the Standard run also exceeds the NO VMR found in POES-FRES, due to more NO being transported down from above. This is also evident at 80 km where the timing of the maximum NO VMR is found in line with the observations in the main phase of the event, although it is largely underestimated. In the aftermath of the event, the Standard NO VMR is higher due to a stronger source region above. Despite the general energy deposition during the quiet period being higher, the POES-FRES underestimates the quiet background level at 70 km. The estimate of the direct production, associated with the first event, is, however, much improved. At 60 km, POES-FRES appears to overestimate the NO VMR compared to the observations. The continued rise of the NO VMR from the first event on DOY 95 to the second event on DOY 105 is consistent with downward transport by the indirect effect from a source region in the mesosphere. Even though the observations also show NO enhancements, the simulated NO appears overestimated.

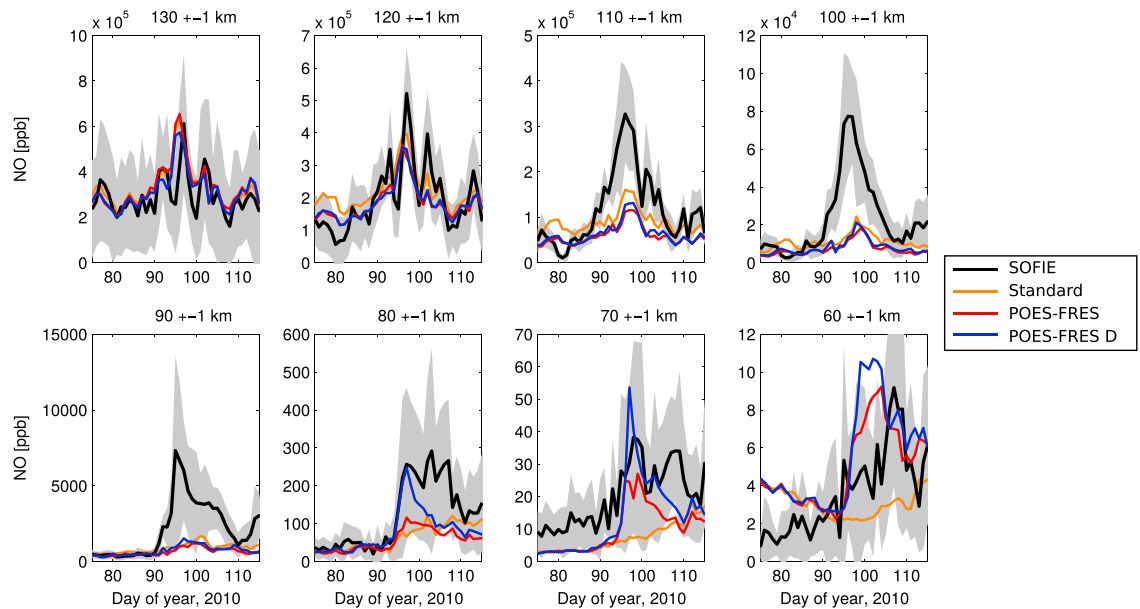

**Figure 5.** Time evolution of NO volume mixing ratio (VMR; ppb) for eight different altitude regions (daily mean, altitude mean ±1 km, and zonal mean of latitude band where Aeronomy of Ice in the Mesosphere-SOFIE is measuring). The black line represents SOFIE observations of NO VMR, where the observational spread is shown as the gray shaded region. The orange line represents NO VMR from the Standard run, the red line is from the POES-FRES run, and the blue line represents from the POES-FRES run with $D$ region chemistry. The two CMEs arrive at days of year (DOYs) 95 and 101. DOY 75 is 16 March, and DOY 120 is 30 April. SOFIE = Solar Occultation For Ice Experiment; POES = Polar Orbiting Environmental Satellites; FRES = full range energy spectrum.

In summary, using the POES-FRES energy deposition lowers the modeled quiet time background level NO VMR at 100- to 120-km altitude compared to the Standard run, and the modeled NO VMR is more in line with the observed NO by SOFIE. During the storm, the inclusion of MEE brings NO at 70–90 km closer to the observations. However, there still remains a strong underestimate at 80–110 km. The underestimate at mesospheric altitudes could be due to missing NO production from ion chemistry reactions that are not included in the standard WACCM chemistry scheme. In the following, we use the new model WACCM-D, which includes detailed ion chemistry in the $D$ region.

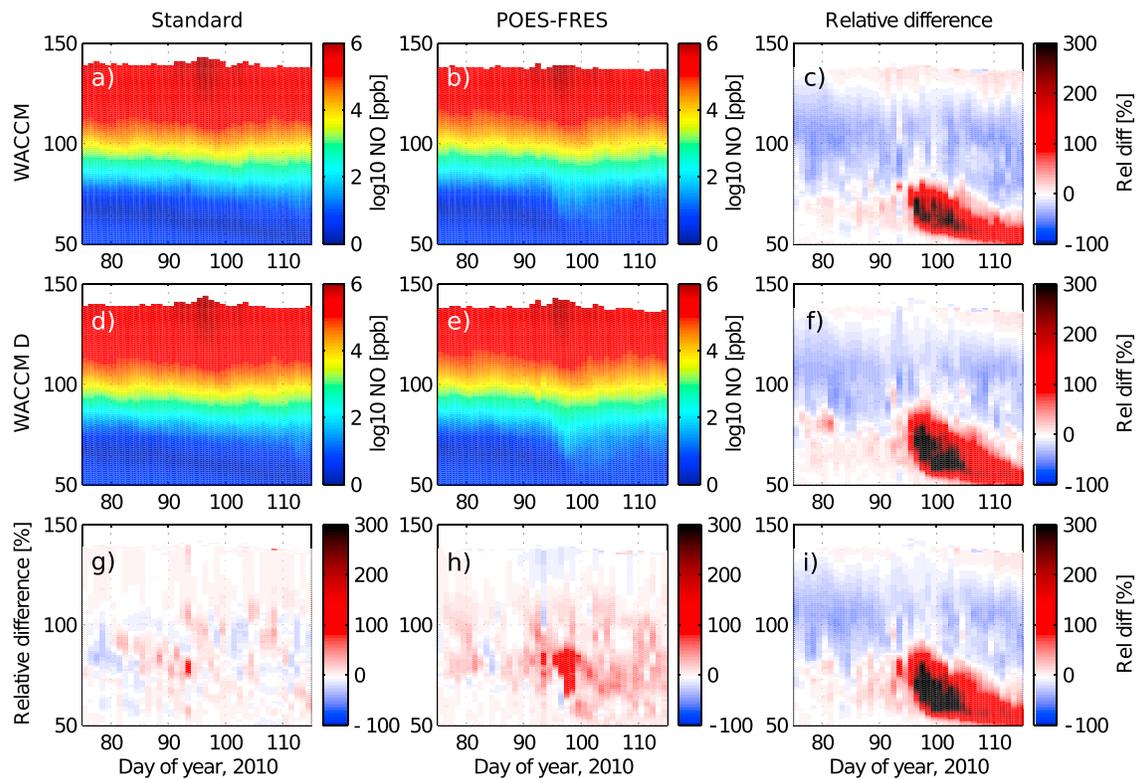

**Figure 6.** Daily mean modeled NO collocated to the Solar Occultation For Ice Experiment observations. Upper panels are standard WACCM, middle panels WACCM-D, and lower panels the relative difference between WACCM and WACCM-D ([WACCM-D-WACCM]/WACCM). The first column is modeled NO with the Standard energy deposition. The middle column is with the POES-FRES energy deposition, and the right column the relative difference ([POES-FRES-Standard]/Standard) in NO volume mixing ratio when using the different energy inputs. Day of year (DOY) 75 is 16 March, and DOY 120 is 30 April. The two coronal mass ejections arrive on 5 and 11 April 2010 (DOYs 95 and 101). WACCM = Whole Atmosphere Community Climate Model; POES = Polar Orbiting Environmental Satellites; FRES full range energy spectrum.

### 4.4. Modeled Nitric Oxide, With *D* Region Ion Chemistry

Finally, we examine the sensitivity of NO production to the choice of chemistry scheme. In WACCM-D, the EEP production of NO can also occur through ion chemistry reactions that are not present in standard WACCM. By using WACCM-D and applying the POES-FRES energy input, we obtain the modeled NO shown as the blue line in Figure 5. The thermospheric NO production from EEP will still occur mainly through reactions with the five ions from the Standard WACCM chemistry, which is confirmed by the same characteristics of NO at and above 90 km. At and below 80 km, in the *D* region, the NO production efficiency is greatly enhanced compared to the Standard run (orange line). In particular, at 80 km, in the main phase of the storm, the same energy input yields a much stronger direct impact in WACCM-D (blue line) than in standard WACCM (red line), and the NO from the POES-FRES D run fits very well with the observations both in timing and increase. The days after the second CME, where most of the NO is associated with the indirect effect, all model runs underestimate the NO at 80 km, due in part due to the underestimation of EEP produced NO at higher altitudes. At 70 km, there is still a pronounced underestimation of the NO level in the geomagnetic quiet period before the storm. The peak associated with the first CME seems to be fairly well in line with observed NO VMR. After DOY 105 the modeled NO VMR drops below the measured NO VMR level. At 60 km in the POES FRES-D run, NO VMR is overestimated during both the prestorm and the main storm phases.

In summary, WACCM-D provides a better estimated NO production at 60–80 km in the main phase of the storm. Comparing the POES-FRES energy input to the Standard run, it is evident that MEEs are important both for direct NO production as well as the indirect effect at and below 80 km. At 90–110 km, however, all model runs with all specifications of electron energy deposition underestimate the storm time NO production. This deficiency propagates down and contributes to an underestimate of NO at 80 km and further below in the aftermath of the events.

### 4.5. Model Sensitivity to Change in Energy Input and Chemistry

The difference between the NO VMR resulting from the Standard and POES-FRES runs (see Figure 6, upper panels) varies throughout April 2010. POES-FRES deposits less energy than the Standard input above 90 km. Above 70 km, both standard WACCM and WACCM-D predict less NO with the POES-FRES energy input, compared to the Standard energy input (see Figures 6c and 6f). The negative difference appears to propagate down from the lower thermosphere and counteracts the extra NO produced by POES-FRES in 90- to 100-km region. This implies that the major difference between the NO VMR in the mesosphere is due to the downward propagation from the lower thermospheric source region.

To fully understand the impact of the added *D* region chemistry, we investigate how the same energy input can result in a different NO VMR in WACCM compared to WACCM-D. Applying just the Standard energy input (see Figure 6, first column) leads to only minor changes in the NO VMR from standard WACCM compared to WACCM-D. No energy is deposited below 90 km (as demonstrated in Figure 1); nevertheless, WACCM-D introduces a general increase in NO in this altitude region at the arrival of the first CME. This appears to be associated with a corresponding weak increase above 90 km, which is propagating down with time. When the second CME arrives, a decrease of up to 20% NO is seen in WACCM-D compared to WACCM. This deficit is seen below 90 km a few days later.

At the arrival of the first CME a strong positive increase up to 100% is found between 70 and 90 km when using the POES-FRES input in WACCM-D compared to the same input in standard WACCM (see Figure 6, middle column). This appears to be increased direct production when using the more complex chemistry in WACCM-D. In the aftermath of the CME, the increase in WACCM-D compared to WACCM also affects the indirect effect on NO as the increase is transported further down in the following days. In the Standard runs, we find both increases and decreases of up to $\sim \pm 20\%$ at 90 km and above.

## 5. Discussion

For a model to reproduce the direct EEP effect on NO, it is essential to have a correct energy deposition rate, as well as an adequate chemical scheme. For the model to reproduce the indirect effect, the two first criteria need to be in place in addition to a realistic, effective descent rate in the MLT region and in the lower mesosphere/upper stratosphere.

In the comparison with SOFIE measurements, both WACCM and WACCM-D with both energy inputs match the NO VMR very well at 120 km and above. Both the quiet time background level of NO, and the timing and strength of the increase during the geomagnetic storm fit well. In the main production region, at 90- to 110-km altitude, the Standard model run overestimates the background level and underestimates the storm time NO increase. As the lower thermospheric NO is the source region for NO in the mesosphere, this shortage of NO propagates downward and affects the entire mesosphere. The direct production in the mesosphere is, however, significantly increased by including MEE from FRES (in the POES-FRES run) and the ion chemistry from WACCM-D (in the POES-FRES D run). In the following we discuss potential sources for the discrepancies. We will assess the different energy inputs, the two different chemical schemes in WACCM, and evaluate the importance of dynamics with respect to the indirect effect.

### 5.1. Energy Input

The energy input in the Standard and POES-FRES is based solely on electron precipitation, which is expected to be the dominating part of the particle precipitation during the April 2010 event. WACCM also includes ionization from protons, based on proton measurement on the Geostationary Operational Environmental Satellite. The Geostationary Operational Environmental Satellite measures protons with energies larger than $\sim 1$ MeV, which deposit their energy below 100 km. Lower-energy protons in the energy range from a few tens of kiloelectron volts to $\sim 1$ MeV might also be important for the NO production above $\sim 100$ km. Looking at the measurements by the MEPED proton fluxes during our event (not shown), we find elevated proton fluxes in the main phase of both CMEs, even though no SPE occurs at this time. Including the associated ionization also from low-energy protons could improve the modeled storm time NO above 100 km.

In the current POES-FRES energy input we have assumed isotropy over the loss cone using the 0° detector on both TED and MEPED. In the case of an anisotropic electron pitch angle distribution, using a single telescope will overestimate or underestimate the electron fluxes being lost in the atmosphere. In general, the auroral electron fluxes, depositing their energy above $\sim 90$ km, are expected to have an isotropic pitch angle distribution, while the higher-energy electrons depositing their energy below ∼90 km are often strongly anisotropic over the loss cone. Nesse Tyssøy et al. (2016) estimate the loss cone fluxes by combining the measurements from both of the MEPED telescopes with electron pitch angle distributions from theory of wave-particle interactions in the magnetosphere. The theories related to the wave-particle pitch angle scattering are not valid at the auroral energies (<30 keV) as the electrons will be accelerated parallel to the magnetic field. Nevertheless, in the search for a potential missing NO production source in WACCM, to what extent the assumption of isotropy over the loss cone holds for the auroral electron energies should be investigated. The medium to relativistic electron energies (>30 keV) often has a highly anisotropic distribution over the loss cone. This implies that the POES-FRES energy input, which is based on the 0° detector, will provide an underestimate of the precipitating fluxes (Nesse Tyssøy et al., 2016; Rodger et al., 2013). The current energy input appears to reproduce fairly well the NO VMR at 80 and 70 km in WACCM-D in the main phase of the first CME event. There is, however, a lack of NO VMR in both WACCM and WACCM-D associated with the second CME and in the aftermath of the event when using POES-FRES. To what extent this relates to an underestimation of the direct NO production or due to the indirect effect caused by the underestimation of NO in the lower thermosphere is yet to be investigated.

MEEs have a significant impact not only at the altitude they deposit their energy but also on the indirect effect due to transport in the subsequent days. The additional NO VMR is increased by 300% and 400% by the POES-FRES energy input compared to the Standard run. This emphasizes the importance of a realistic altitude profile of the electron energy deposition in order to predict both the timing and strength of both the direct and indirect effect.

### 5.2. Chemistry

With the ion chemistry used in WACCM-D instead of the effective production rate, Andersson et al. (2016) show an increase of 30–130% in $NO_x$ at 75–80 km compared to standard WACCM, which leads to a better agreement with the observations during the January 2005 SPE. We also find an increase of up to 100% in the same altitude range during the EEP event in April 2010. Using POES-FRES as energy input, we see that the added electron energy deposition below 90 km enhances both the NO direct and indirect effect. Applying the $D$ region chemistry there is hence a better agreement with the observations in respect to the direct impact in the main phase of the event. There persists, however, a significant NO deficit at 90- to 110-km altitude compared to the observations, even in WACCM-D, indicating inadequate production rates in lower thermosphere.

Rusch et al. (1981) and Sinnhuber et al. (2012) point out that there are huge differences in the final results of model computations of NO enhancements from particle precipitation that depend strongly on the branching ratios of the N atoms produced. The $N(^4S)$ can contribute to both production and loss of NO in the following processes:

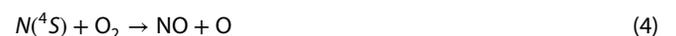
$$N(^4S) + O_2 \rightarrow NO + O \tag{4}$$

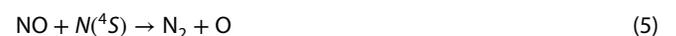
$$NO + N(^4S) \rightarrow N_2 + O \tag{5}$$

where equation (5) is highly temperature dependent. Nieder et al. (2014) illustrate how the effective production rate of $NO_x$ due to EEP varies with density, temperature, and abundance of atmospheric constituents and trace gasses using the University of Bremen Ion Chemistry model. They show that the simplified NO production relation from equation (3) underestimates the $NO_x$ production above ∼80-km altitude, and our results are consistent with that.

We note that there is an additional production of NO in WACCM-D, of the order of 100% in presence of MEE and of the order of 10% in absence of MEE. Although WACCM-D was tailored for the $D$ region (below 90 km), it is applied in the whole vertical domain in these simulations. Since the abundance of ion clusters and negative ions decreases sharply above the $D$ region, the complex WACCM-D chemistry mostly reduces to the simple five-positive ion chemistry in the standard WACCM in the thermosphere. In the $D$ region where the ion clusters are more abundant, multiple ion-ion recombination reactions can contribute to produce NO (Verronen et al., 2016). In addition, there is a number of positive ion reactions that can produce NO, for example, when water cluster ions based on $NO^+$ react with $HO_x$ or $H_2O$ to convert into hydrated water ion clusters. These results highlight the importance of MEE to enhance NO production, and it remains to be determined if there is missing ion cluster chemistry in the $E$ region (above 90 km) where we still observe a deficit compared to the SOFIE observations. Also, the reaction rates of the existing $E$ region chemistry could be incorrect, as reaction rates will change with temperature, density, and abundance of different gasses.

Both the branching ratio of $N(^4S)$ and $N(^2D)$ in the thermosphere and some missing ions and inaccurate reaction rates could contribute to the underestimate of NO production in 90–110 km during the EEP events studied. The transition from energy-dependent branching ratios to constant ratios occurs around 200 km (Porter et al., 1976), so this should not play a role in the altitude range we investigate.

### 5.3. Dynamics

Meraner and Schmidt (2016) analyzed the role of advection, molecular diffusion, and eddy diffusion for the transport of nitrogen oxides through the mesopause region in the Northern Hemisphere, by using simulations of the general circulation and chemistry model, Hamburg Model of Neutral and Ionized Atmosphere. Their results clarify which dynamical conditions favor the intrusion of thermospheric air to the mesosphere and indicate how potential trends in eddy diffusion may change the transport characteristics and found that molecular diffusion and advection are the dominant processes for the transport of $NO_x$ in the lower thermosphere. Orsolini et al. (2017) examined the role of eddy diffusion on the NO distribution and dynamical conditions of strong thermospheric to mesospheric transport in the Northern Hemisphere and found missing advection during dynamical conditions of strong thermospheric to mesospheric transport in the Northern Hemisphere.

WACCM has a relatively weak vertical transport across the mesopause layer (Smith et al., 2011), but the enhanced eddy diffusion allows more NO to be captured and transported down by the mean meridional circulation. The altitude of the reversal from downward to upward transport over the pole might be located at the wrong altitude in WACCM, due to uncertainties in the parameterization of the gravity waves driving the mean meridional circulation. This would lead to incorrect effective descent rates in the model. However, Hendrickx et al. (2018) compared the downward transport of NO through the mesopause in SOFIE and WACCM via an epoch analysis and found similar descent rates in both data sets.

The indirect effect we observe in the modeled NO likely contributes to the discrepancies seen in the model NO in the lower mesosphere; hence, the remaining deficit in NO near 100 km has an impact also at lower altitudes. The deficit found in the mesosphere and below is a combination of both a lack of electron energy deposition input and the production deficit above. Future studies should further investigate deficiencies and uncertainties in the chemistry, either photochemistry or ion cluster chemistry, in particle forcing and in transport across the mesopause region.

## 6. Conclusions

The comparison between NO modeled by WACCM and NO measured by AIM satellite's SOFIE instrument during the April 2010 event highlights the importance of the MEE direct effect and of ion chemistry. The timing of the event is well modeled by WACCM, and so is the NO abundance above 120 km. WACCM underestimates the storm time NO abundance in the main source region (90–110 km), and the quiet time NO in the thermosphere (100–120 km) is higher in the model than in the observations. These discrepancies remain also when using improved energy input and a more complex ion chemistry, indicating that something else is missing. The deficiency propagates down, contributing to an underestimate of NO at 90 km and below in the aftermath of the storm.

At and below 80 km including MEE is important both for direct NO production and the indirect effect. By using WACCM-D the direct NO production in this region is increased further. We conclude that $D$ region chemistry, dynamics and MEE are important for the model to predict realistic NO levels in the mesosphere, and that some source of extra production is needed for WACCM to fully reproduce the observed NO at 90–110 km.

**Acknowledgments**
The research has been funded by the Norwegian Research Council, through project 222390, "Solar-Terrestrial Coupling through High Energy Particle Precipitation in the Atmosphere: A Norwegian contribution." D. R. M. is supported in part by NASA LWS grant NNX14AH54G. The National Center for Atmospheric Research is sponsored by the U.S. National Science Foundation. H. N. T. and Y. O. were supported by the Norwegian Research Council under contract 223252/F50. We thank NOAA's National Geophysical Data Center for providing the NOAA POES data. Model runs are available at the Norwegian Centre for Research data (nsd.no). The AIM-SOFIE data are found online at http://sofie.gats-inc.com/sofie/index.php.